\documentclass[%
reprint,
amsmath,amssymb,
aps,
floatfix,
]{revtex4-1}

\usepackage{graphicx}
\usepackage{dcolumn}
\usepackage{amsmath}
\usepackage{amssymb}
\usepackage{bm}
\usepackage[bbgreekl]{mathbbol}
\usepackage{color} 
\usepackage{braket}
\usepackage{gensymb}
\usepackage[dvipsnames]{xcolor}
\usepackage{lpic}

\usepackage{xcolor}
\usepackage[percent]{overpic}
\usepackage{stackengine}
\usepackage{physics}
\usepackage{siunitx}
\usepackage{textcomp}
\usepackage{slashbox}
\usepackage{lipsum}
 \usepackage{hyperref}
 \hypersetup{
    colorlinks=true,
    linkcolor=blue,
    filecolor=magenta,
    citecolor=blue,
    urlcolor=cyan,
}

\usepackage[mathscr]{eucal}

\usepackage{amsfonts}
\usepackage[colorinlistoftodos]{todonotes}
\usepackage{mathtools}
\usepackage{empheq}
\usepackage{esvect}
\usepackage{rotating}
\usepackage{ulem} 

\DeclareSymbolFontAlphabet{\mathbbm}{bbold}
\DeclareSymbolFontAlphabet{\mathbb}{AMSb}

\begin{document}

\preprint{APS/123-QED}

\title{Continuous Variable Entanglement in an Optical Parametric Oscillator Based on a Nondegenerate Four Wave Mixing Process in Hot Alkali Atoms}
\author{A. Monta\~na Guerrero$^1$}
 \email{amontanag@usp.br}
\author{R. L. Rinc\'on Celis $^1$}
\author{P. Nussenzveig$^1$}
\author{M. Martinelli$^1$}
\author{A. M. Marino$^2$}
\author{H. M. Florez$^1$}
 \email{hans@if.usp.br}
\affiliation{$^1$ Instituto de F\'{\i}sica, Universidade de S\~ao Paulo, 05315-970 S\~ao Paulo, SP-Brazil}
\affiliation{$^2$ Center for Quantum Research and Technology and Homer L. Dodge Department of Physics and Astronomy, The University of Oklahoma, Norman, Oklahoma 73019, USA}

\date{\today}

\begin{abstract}
We present the measurement of entanglement between twin beams generated with a doubly resonant optical parameter oscillator (OPO) based on four-wave mixing in hot $^{85}$Rb vapor above threshold.
This is the first measurement of entanglement in an OPO with $\chi^{(3)}$ media above threshold.
We reconstruct the covariance matrix for several configurations and with a full picture of the four side band mode state, we study entanglement between all possible bi-partitions.
We show a robust generation of entanglement with stronger generation for a specific pair of modes.
For this system, we show that atomic density is a determinant factor for generation and loss of quantum correlations.
The generation of entangled fields by an atomic OPO close to atomic resonance of alkali atoms enables natural integration into quantum networks.


\end{abstract}

\pacs{Valid PACS appear here}
\maketitle

Entanglement, an important quantum phenomena, is a keystone for the development of a quantum network usable in quantum communication and quantum computation  \cite{Nielsen11}.
Among the diversity of quantum entanglement sources, the ones in the continuous variable domain have shown great potential for applications and versatility in the generation of entangled states.
Specifically, solid-state optical parametric oscillators (OPOs) have been used to generate correlated and entangled light beams below \cite{Ou92} and above \cite{Villar05} the OPO oscillation threshold based on $\chi^{(2)}$ media.
OPOs also have the capability of generation of large ensembles of multimode entangled fields ~\cite{Pysher11,Pinel12,Shota13,Moran14}. 
Furthermore, OPOs based on optical chips have been engineered to generate quantum correlated light based on the $\chi^{(3)}$ process of four wave mixing (FWM) \cite{Avik15} below the oscillation threshold.

Quantum correlated light can also be generated through a free space FWM process in hot atomic vapors \cite{PDLett07,Marino08}.
These systems present higher gains than their solid-state counterpart and the absence of phonon noise, hence a higher efficiency and stronger correlations both in amplitude and phase quadratures are expected.
In a recent  work  \cite{Guerrero20}, we demonstrated the generation of an intensity difference squeezing of -2.7 dB from twin beams generated by an OPO above the oscillation threshold in atomic vapor media.

In this Letter, we demonstrate the first measurement (to the best of our knowledge)  of entanglement between twin beams generated by a doubly resonant OPO operating above threshold using a $\chi ^{(3)}$ media.
The amplifying process comes from a FWM process in a hot vapor cell of alkali atoms within a cavity.
The quantum correlations of the system are tested using the Duan \textit{et al.} \cite{Duan00} criterion, where we compute the addition and difference of the amplitude $\hat{p}$ and phase $\hat{q}$ related quadratures  $\hat{p}_-=(\hat{p}_a-\hat{p}_b)/\sqrt{2}$ and $\hat{q}_+=(\hat{q}_a+\hat{q}_b)/\sqrt{2}$ operators, for modes $a$ and $b$, such that a violation on the inequality
\begin{eqnarray}\label{eq:Duan}
   \Delta ^2 \hat{p}_-+\Delta ^2 \hat{q}_+ \geq 2 ,
\end{eqnarray}
demonstrates a continuous variable entangled state.
This corresponds to the quantum correlations involving the two intense output beams generated by the OPO, considering a given analysis frequency. 
A more detailed analysis can be done once we discriminate each of the involved sidebands of these two intense fields.
The detailed analysis of the structure of entanglement among the four modes involved uses the Simon version of Positivity under Partial Transposition (PPT) \cite{Simon00} criteria, from the full reconstruction of the covariance matrix of the four side band modes of the twin beams \cite{Barbosa13a}. 
\begin{figure}[h!]
\begin{overpic}[width=86mm]{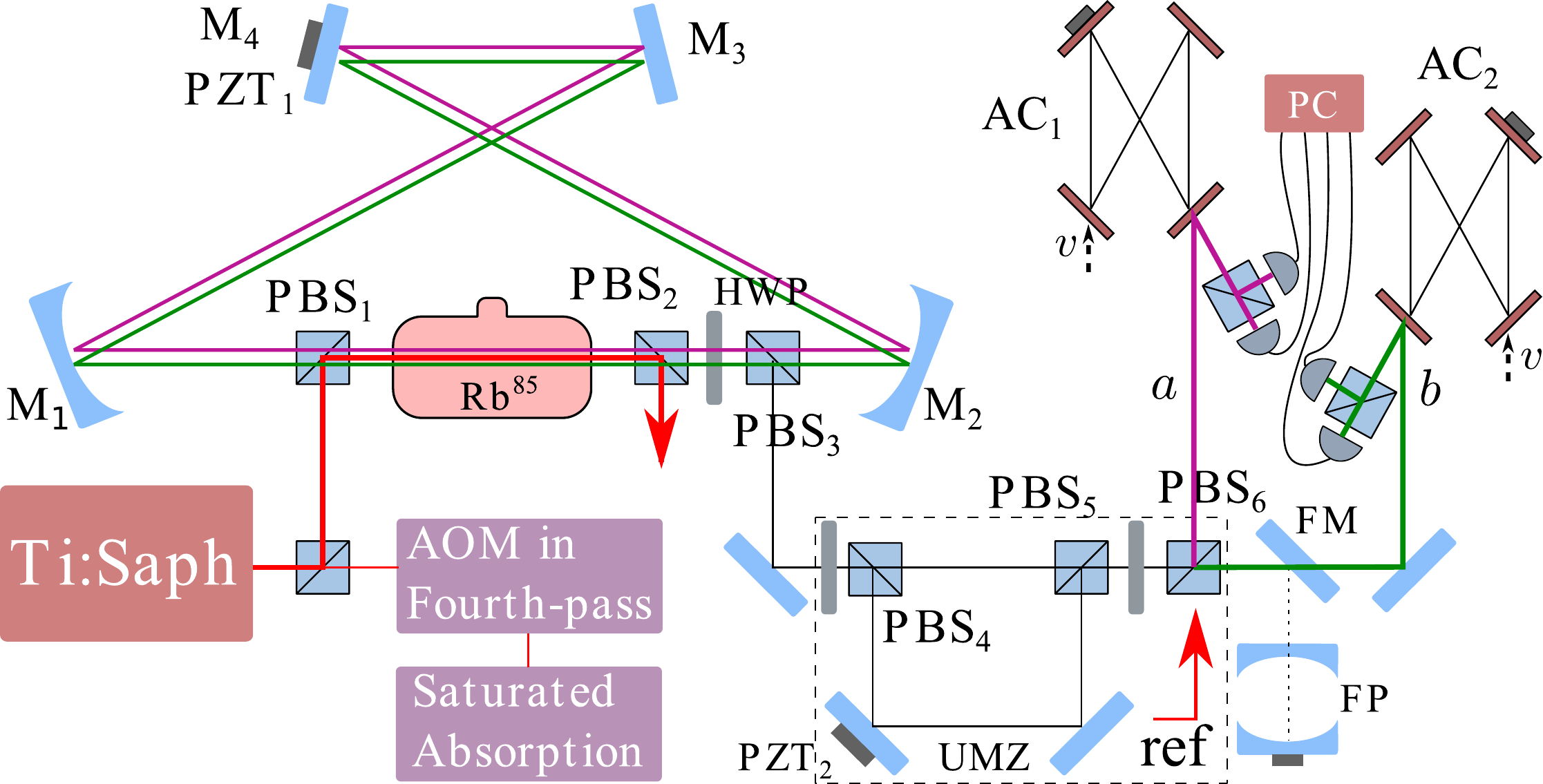}
\put(0,45){(a)}
\end{overpic}
\begin{overpic}[width=86mm]{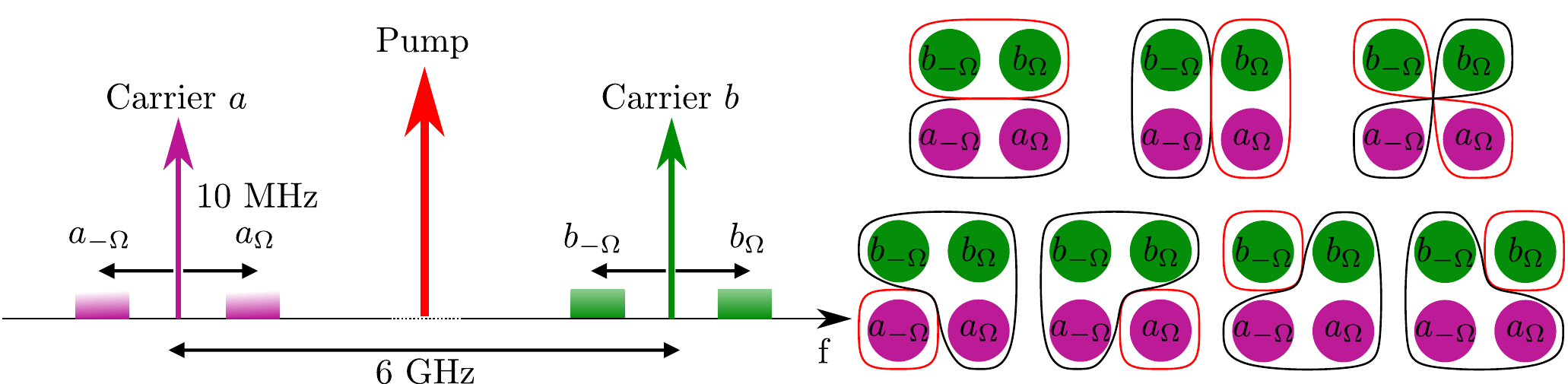}
\put(0,20){(b)}
\end{overpic}
\caption{(a) Sketch of the experimental setup. PBS: polarising beamsplitter cube, HWP: half-wave plate, UMZ: unbalanced Mach-Zehnder interferometer, M: mirror, FP: confocal Fabry-Perot,  FP: flip-mirror, AC: analysis cavity and PC: demodulating chain and data acquisition. (b) Representation of the sidebands of mode carriers ``$a$" and ``$b$" and all seven possible bipartitions.}
\label{fig:Setup}
\end{figure}
The experimental setup is shown in Fig.~\ref{fig:Setup}(a) with a detailed description presented in the Supplementary Material~\cite{suppl}. 
The OPO consists on a bow-tie cavity with high reflectivity mirrors and a free spectral range (FSR) of 404.7(3) MHz, allowing a doubly resonant operation between a pair of beams separated by a frequency of 6 GHz, twice the ground state hyperfine atomic splitting of $^{85}$Rb.
A $^{85}$Rb vapor cell with anti-reflection coatings is placed inside the cavity and is heated to 97$^\circ$C for high optical density, as needed for an efficient FWM process.

The pump beam is colinear with the cavity mode and is injected by a polarising beam splitter PBS$_1$ and removed by  PBS$_2$.
It is generated by a Ti:sapphire laser tuned to the D1 (@795nm) line of $^{85}$Rb being locked $\Delta_1=$0.82 GHz to the blue of the $5^2S_{1/2}F=2\rightarrow 5^2P_{1/2}F=3$ transition.
A half wave-plate (HWP) and PBS$_3$ are used in order to control the output coupling of the cavity, with a finesse that ranges from 5 to 30 for a field far from the atomic resonance.
We use an unbalanced Mach-Zehnder (UMZ) interferometer~\cite{Huntington05} to separate the twin beams with the same output polarization with a efficiency of 98\%.

After spatial separation, we use the technique of Resonator Detection (RD) \cite{Barbosa13a} to reconstruct the full covariance matrix.
The RD technique consists on two analysis cavities, one for each beam, that interfere the incoming field with the intra-cavity field.
As a consequence of the dispersive properties of an optical resonator close to resonance, the cavity leads to a discrimination of the sideband modes before photodetection, enabling a full reconstruction of the covariance matrix.
Following balanced detection for each cavity, the photocurrent is splitted in two, demodulated using in-quadrature signals at a chosen analysis frequency and processed in a computer.
The whole system has an overall detection efficiency of 91\%, accounting for optical losses and photodetector quantum efficiency.

Assuming a Gaussian quantum state, the full state can be described by the covariance matrix
\begin{eqnarray}
\mathbf{V}=\frac{1}{2}\big(\langle \Vec{X}\cdot \Vec{X}^T \rangle+\langle \Vec{X} \cdot \Vec{X}^T \rangle ^T\big),
\end{eqnarray}
where for our case $\Vec{X}=(\Vec{X_\textup{s}},\Vec{X_\textup{a}})^T$ is the column vector of organized symetric $\Vec{X_\textup{s}}=(p_{1\textup{s}},q_{1\textup{s}},p_{2\textup{s}},q_{2\textup{s}})$ and antisymetric $\Vec{X_\textup{a}}=(p_{1\textup{a}},q_{1\textup{a}},p_{2\textup{a}},q_{2\textup{a}})$ amplitude  $p$ and phase $q$ quadratures of the two beams.
These operators are linear combinations of symmetric and antisymmetric side band modes, $p_{(\textup{s},\textup{a})}=(p_{\Omega} \pm p_{-\Omega})/\sqrt{2}$ and  $q_{(\textup{s},\textup{a})}=(q_{\Omega} \pm q_{-\Omega})/\sqrt{2}$, where $p_{\Omega}=(a_{\Omega}+a^{\dagger}_{\Omega})$ and $q_{\Omega}=-i(a_{\Omega}-a^{\dagger}_{\Omega})$ at frequency $\Omega$ are linear combinations of field operators on sideband modes $a_{\pm\Omega}$ ($b_{\pm\Omega}$) as depicted in Fig.~\ref{fig:Setup}(b).

The symmetric covariance matrix for the two beams $V_{(\textup{s}/\textup{a})}^{(12)}$ is
\begin{eqnarray}
V_{(\textup{s}/\textup{a})}^{(12)}=
\begin{pmatrix}
V_\textup{s}^{(12)} & C_{(\textup{s}/\textup{a})}^{(12)}\\
\big(C_{(\textup{s}/\textup{a})}^{(12)}\big)^T & V_\textup{a}^{(12)}
\end{pmatrix},
\end{eqnarray}
\begin{eqnarray}\nonumber
V_\textup{s}^{(12)}=
\begin{pmatrix}
\alpha^1 & \gamma^1 & \mu & \epsilon \\
\gamma^1 & \beta^1 & \xi & \nu \\
\mu & \xi & \alpha^2 & \gamma^2 \\
\epsilon & \nu & \gamma^2 & \beta^2
\end{pmatrix},
C_{(\textup{s}/\textup{a})}^{(12)}=
\begin{pmatrix}
\delta^1 & 0 & \kappa & -\eta \\
0 & \delta^1 & \tau & -\lambda \\
-\lambda & \eta & \delta^2 & 0 \\
-\tau & \kappa & 0 & \delta^2
\end{pmatrix},
\end{eqnarray}
\begin{eqnarray}\nonumber
V_\textup{a}^{(12)}=
\begin{pmatrix}
\beta^1 & -\gamma^1 & \nu & -\xi \\
-\gamma^1 & \alpha^1 & -\epsilon & \mu \\
\nu & -\epsilon & \beta^2 & -\gamma^2 \\
-\xi & \mu & -\gamma^2 & \alpha^2
\end{pmatrix},
\end{eqnarray}
where parameters $\alpha, \beta...,\mu, \eta...$ correspond to the full variances and correlations needed for the reconstruction of the state with the RD technique.
The photocurrent spectral noise power of RD for each individual beam is  $S_{RD}=\langle I_\Omega I_{-\Omega} \rangle$ that results in
\begin{eqnarray}\label{eq:variance}
    S_{RD}(\Delta)&=&c_\alpha \alpha+c_\beta \beta+c_\gamma \gamma +c_ \delta \delta +c_v,
\end{eqnarray}
where $c_\alpha=\big| g_+ \big|^2$, $c_\beta=\big| g_-\big|^2$, $c_\gamma+i c_\delta = 2g^\ast_+ g_-$ and $c_v=1-c_\alpha-c_\beta$ are functions that depend only on cavity reflection and transmission coefficients by the relations
\begin{eqnarray}
g_+=\frac{1}{2} \bigg( \frac{r^*(\Delta)}{|r(\Delta)|}r(\Delta+\Omega)
+ \frac{r(\Delta)}{|r(\Delta)|}r^*(\Delta-\Omega) \bigg),\\
g_-=\frac{i}{2} \bigg( \frac{r^*(\Delta)}{|r(\Delta)|}r(\Delta+\Omega)
- \frac{r(\Delta)}{|r(\Delta)|}r^*(\Delta-\Omega) \bigg).
\end{eqnarray}

The reflection $r(\Delta)=-(\sqrt{d}+2i\Delta)/(1-2i\Delta)$ is calculated under the assumption of high finesse of the analysis cavity  with $\Delta=(\omega-\omega_c)/\delta \nu_{ac}$ being the detuning between the longitudinal mode frequency $\omega$ and the resonator frequency $\omega_c$ normalized to the resonator bandwidth $\delta \nu_{ac}$ and with a fraction of reflected light at exact resonance $d=|r(0)|^2$.
For analysis frequencies larger than $\sqrt{2} \delta \nu_{ac}$ we have a complete conversion of the incident phase fluctuations into amplitude fluctuations of the reflected beam.
When this condition is satisfied, at values $\Delta=\pm\delta\nu_{ac}/2$, $c_\alpha \approx 0$ and $c_\beta \approx 1$ we have access to the phase fluctuations $\beta=\Delta ^2 q_\textup{s}$.
For the opposite case, when $\Delta=0$ and for $\abs{\Delta} \gg \delta\nu_{ac}$, $c_\alpha \approx 1$ and $c_\beta \approx 0$ we recover amplitude fluctuations $\alpha=\Delta ^2 p_\textup{s}$.
Correlations involving the side band modes of one beam are expressed by $\langle p_\textup{s} q_\textup{s} \rangle=\gamma$ and $\langle p_\textup{s} p_\textup{a} \rangle=\delta$.
\begin{figure}[h!]
    \centering
\begin{overpic}[width=86mm]{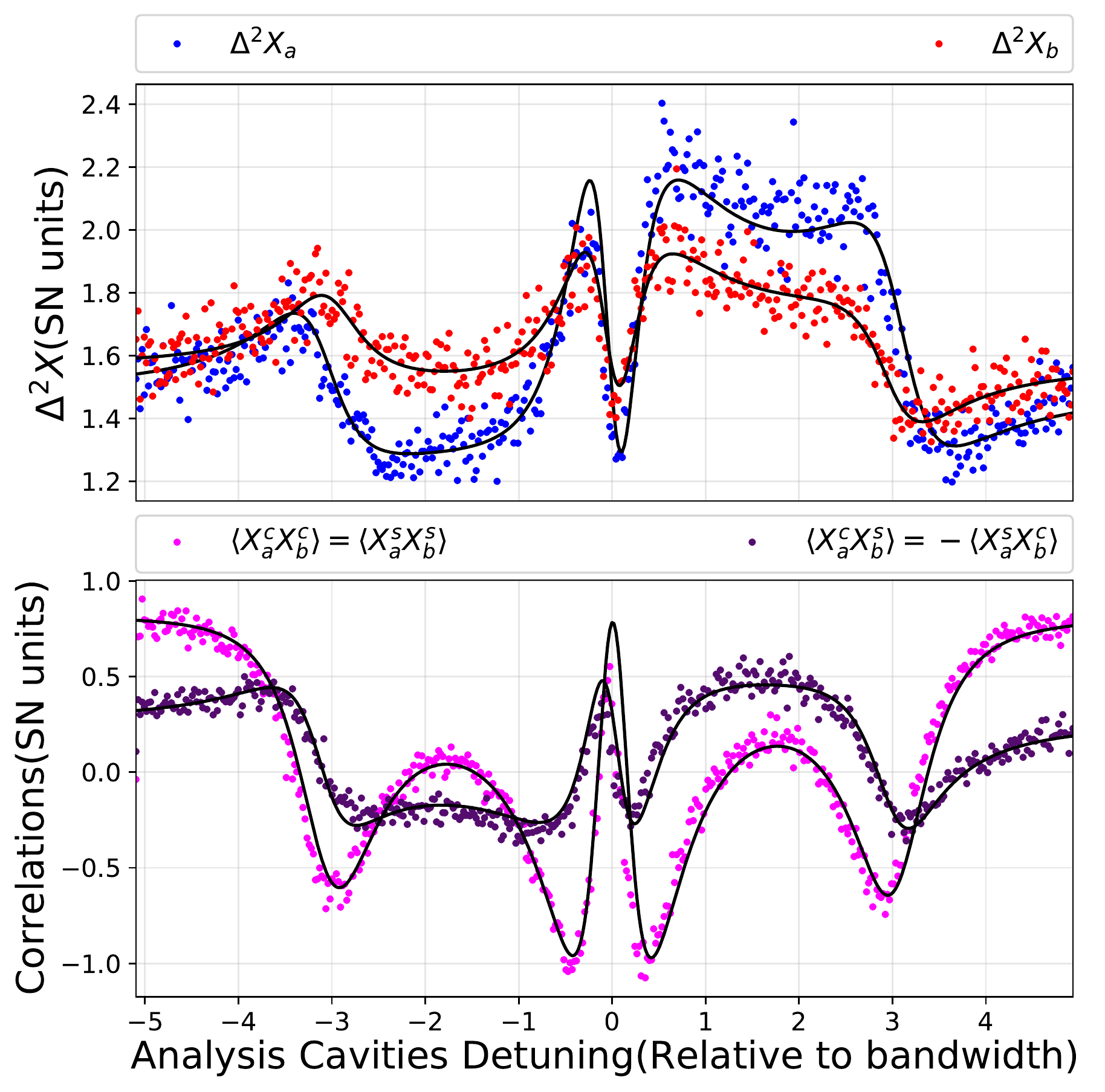}
\put(14,56){(i)}
\put(14,11){(ii)}
\end{overpic}
\caption{Variance (i) and correlations (ii) of the generalized quadrature $X_i$ normalized to the shot-noise level (SN) for the output beams of the OPO, modes ``$a$" and ``$b$". T=91$^{\circ}$C, $\Delta_1=0.82$ GHz, $\mathcal{F}=15$, $\sigma=1.42$, $P_{Th}$=199 mW, $\Omega=$10MHz. Curves correspond to a non-linear fit.
Superscripts c and s represent cosine and sine demodulation of photocurrents.}
    \label{fig:Var_Corr_individual}
\end{figure}

The photocurrent spectral noise is recorded as a function of the synchronous analysis cavity resonant frequency sweep in Fig.~\ref{fig:Var_Corr_individual}(i) for beams ``$a$" and ``$b$" generated by the OPO.   
The resonator bandwidth of our analysis cavities is $\delta\nu_{ac}=3.2$ MHz, therefore, for  analysis frequencies higher than 4.6 MHz, we will have a complete conversion of phase fluctuations into amplitude fluctuations, as is the case of 7 and 10 MHz analysis frequencies that are studied here.
The correlations between two modes are given by $S_{RD}=\big\langle I^{(1)}_\Omega, I^{2)}_{-\Omega} \big\rangle$, expression that can be reorganized into
\begin{eqnarray}\label{eq:correlationsRe}
\Re{\big\langle I^{(1)}_\Omega I^{(2)}_{-\Omega}\big\rangle}=c_\mu \mu +c_\eta \eta +c_\epsilon \epsilon+c_\kappa \kappa \nonumber \\
+c_\xi \xi +c_\lambda \lambda+c_\nu \nu+c_\tau \tau,\\
\Im{\big\langle I^{(1)}_\Omega I^{(2)}_{-\Omega}\big\rangle}=c_\mu \eta-c_\eta \mu +c_\epsilon \kappa-c_\kappa \epsilon \nonumber \\\label{eq:correlationsIm}
+c_\xi \lambda -c_\lambda \xi+c_\nu \tau-c_\tau \nu,
\end{eqnarray}
where $c_\mu+i c_\eta = g^\ast_{+_1} g_{+_2}$, $c_\epsilon+i c_\kappa = g^\ast_{+_1} g_{-_2}$, $c_\xi+ c_\lambda = g^\ast_{-_1} g_{+_2}$ and $c_\nu+i c_\tau = g^\ast_{-_1} g_{-_2}$.
We see that $\Re{\big\langle I^{(1)}_\Omega I^{(2)}_{-\Omega}\big\rangle}$ and $\Im{\big\langle I^{(1)}_\Omega I^{(2)}_{-\Omega}\big\rangle}$ depend on the same unknown parameters so this redundacy improves experimental precision.
We see in Fig.~\ref{fig:Var_Corr_individual}(ii) the real and imaginary parts of $\big\langle I^{(1)}_\Omega, I^{2)}_{-\Omega} \big\rangle$.
Notice that all $c_i$ functions for all $\alpha, \beta...,\mu, \eta...$ depend only on the analysis cavities parameters.
We perform three stages of measurement that are explained in \cite{suppl} and use the package LMFIT (Non-Linear Least-Squares Minimization and Curve-Fitting for Python) to fit the parameters of Eqs. (\ref{eq:variance}),  (\ref{eq:correlationsRe}) and (\ref{eq:correlationsIm}) as shown in Figs.~\ref{fig:Var_Corr_individual} (i) and (ii), respectively.
\begin{figure}[h!]
    \includegraphics[width=86mm]{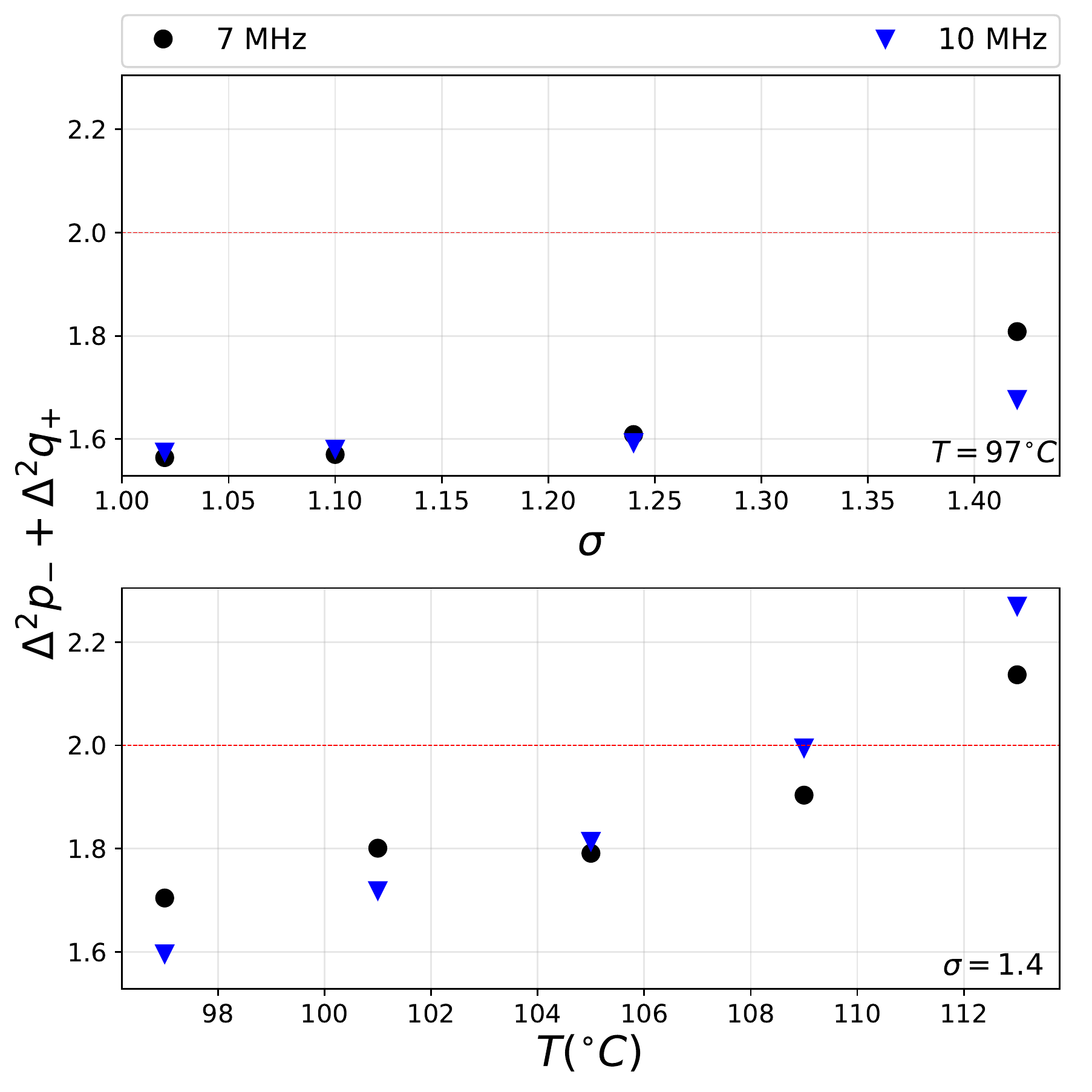}
    \caption{Duan analysis as a function of $\sigma=P/P_{Th}$ with $P_{Th}$=199 mW (top).
    Duan analysis as a function of temperature, $P_{Th}$=174, 125, 90, 61.8, 50 mW for T=97, 101, 105, 109, 113 $^\circ$C respectively (bottom).
    Analysis frequency of 7 MHz and 10 MHz with $\mathcal{F}=15$. For values smaller that  2 we have demonstration of entanglement.}
    \label{fig:Duan_Sigma_Temperatura}
\end{figure}

From the full covariance matrix in the symmetric and antisymmetric basis, which are superposition of the sidebands, we apply the Duan criteria \cite{Duan00}.
We compute the variances in Eq. \ref{eq:Duan} for analysis frequencies of 7 and 10 MHz as a function of the pump power normalized to the oscillation threshold pump $\sigma=P/P_{Th}$ for $T=97^{\circ}C$, as depicted in Fig.~\ref{fig:Duan_Sigma_Temperatura} (top). 
In order to demonstrate a violation of Eq. \ref{eq:Duan}, we compute  $\Delta ^2 \hat{p}_-$ and $\Delta ^2 \hat{q}_+$ separately, with a minimum value of $\Delta^2 p_-=0.71(14)$ and $\Delta^2 q_+=0.85(14)$, showing amplitude difference squeezing as was shown in \cite{Guerrero20} and demonstrating phase sum squeezing for the first time in this kind of OPO.
We show a maximum violation in the Duan criteria of $\Delta^2 p_-+\Delta^2 q_+ =1.56 (14)\ngeq 2$. 

Unlike a solid state OPO, in which the phonon noise degrades the squeezing in the phase quadrature \cite{Coelho09}, atoms do not introduce such excess of noise.
The violation remains robust as we increase the pump power, and the slight reduction in the violation of Duan inequality is compatible with the coupling to the pump mode in the doubly resonant OPO, as described in \cite{Ribeiro20}.
On the other hand, since the analysis frequencies are smaller than the OPO bandwidth of 27 MHz for this configuration, we do not see a clear dependency on this parameter.

This makes the $\chi^{(3)}$ OPO based on atomic media a good source for high level of entanglement above threshold operation, being more robust against the degradation of phase correlations.
Nevertheless, this advantage should be treated with care. 
We also compute the Duan criteria as a function of temperature in Fig.~\ref{fig:Duan_Sigma_Temperatura} (bottom) showing that entanglement is degraded until the violation is lost for values higher than $109^{\circ}$C.
This is consistent with our previous work \cite{Guerrero20} were intensity squeezing is degraded by temperature increase and eventually is turned into excess of noise.
Therefore, there is an optimal condition, where we have enough atoms for fairly low oscillation threshold thanks to the high gain, but avoiding a dense, temperature broadened sample, where incoherent effects can lead to a degradation on the quantum correlations.
It is curious to notice that in the present case, degradation is observed in both amplitude and phase quadratures.
\begin{figure}[h!]
    \centering
    \includegraphics[width=86mm]{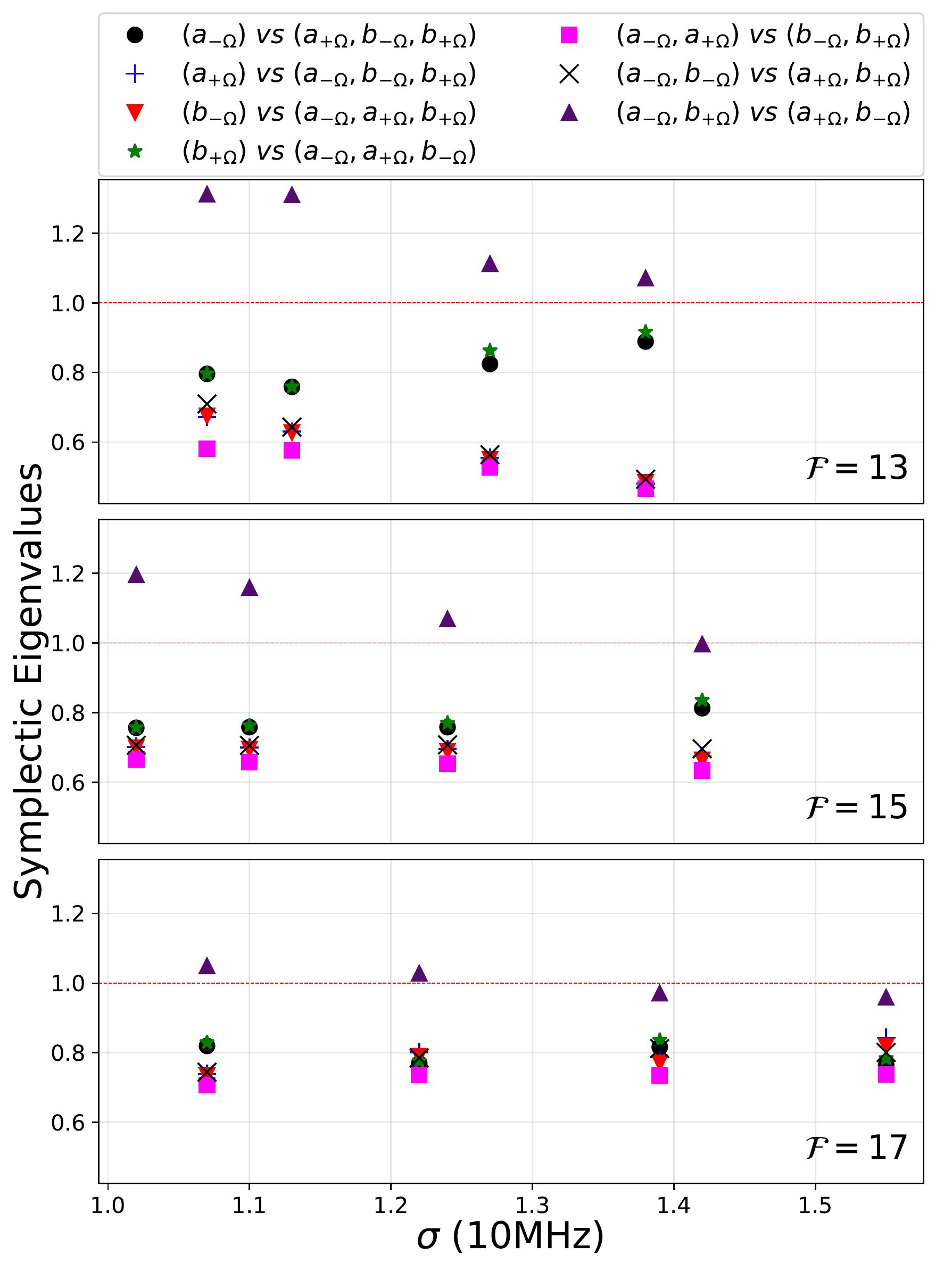}
    \caption{Symplectic eigenvalues after partial transposition, as a function of pump power normalized to the threshold power ($\sigma$) for all possible bi-partitions. Analysis frequency of 10 MHz, temperature 97$^{\circ}$C, $P_{Th}$=274, 199, 138 mW for $\mathcal{F}$=13, 15, 17 respectively.}
    \label{fig:Simplectic_Sigma}
\end{figure}
A richer structure of entanglement is observed if we change the covariance matrix from the symmetric/antisymmetric basis into the side band basis, described by modes   ($a_{-\Omega}$, $a_{+\Omega}$, $b_{-\Omega}$, $b_{+\Omega}$).  
We can apply now the Simon criteria \cite{Simon00}, computing the symplectic eigenvalues of our matrix after a partial transposition for all seven possible bipartitions, as shown in Fig~\ref{fig:Setup}(b).
First, we compute the Simon criteria  as a function of $\sigma$ for three different values of finesse ($\mathcal{F}$=13,15,17), with analysis frequency of 10 MHz and a constant temperature of 97$^{\circ}$C as depicted in Fig~\ref{fig:Simplectic_Sigma}.
Values smaller than one reveals an entangled bipartition.
The particular bipartition ($a_{-\Omega}$, $a_{+\Omega}$) vs ($b_{-\Omega}$, $b_{+\Omega}$) is entangled as expected from the Duan analysis between modes ``$a$" and ``$b$", and presents the maximal violation.
It is interesting to notice that the other $2\times 2$ bipartitions have a quite different behavior.
While ($a_{-\Omega}, b_{-\Omega}$)  Vs ($a_{+\Omega}, b_{+\Omega}$), involving the correlation of the lower against the upper sidebands presents a violation almost as strong as the case of the correlations among beams, the bipartition ($a_{-\Omega}, b_{+\Omega}$) Vs ($a_{+\Omega}, b_{_\Omega}$) is compatible with a separable state.
\begin{figure}[h!]
    \centering
    \includegraphics[width=86mm]{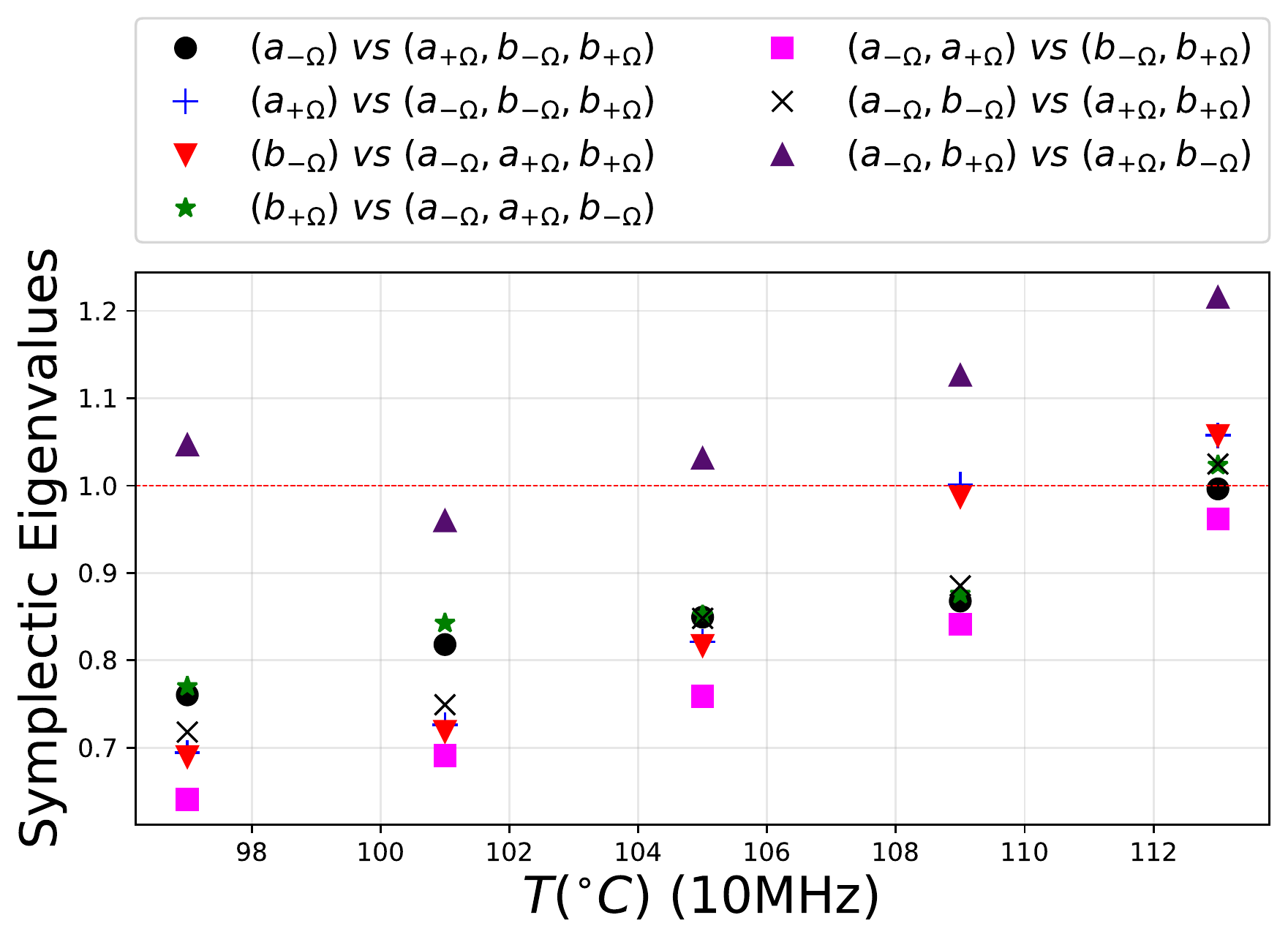}
    \caption{Symplectic eigenvalues after partial transposition, as a function of temperature for all possible bi-partitions. Analysis frequency 10 MHz, $\sigma=1.42$. $P_{Th}$=174, 125, 90, 61.8, 50 mW for T=97, 101, 105, 109, 113 $^\circ$C, respectively}
    \label{fig:Simplectic_Temperaturas}
\end{figure}
That is consistent with the picture for the sidebands in which we have a pair of two mode squeezing operators, acting on the modes $(a_{-\Omega}, b_{+\Omega})$ and $(a_{+\Omega}, b_{-\Omega})$ independently. 
The same dynamics will lead to entanglement of a single sideband against the remaining set of modes.
Oddly, in an open cavity  ($\mathcal{F}$=13), entanglement is not symmetric.
We see that internal bands $a_{+\Omega}$ and $b_{-\Omega}$ are more entangled than $a_{-\Omega}$ and $b_{+\Omega}$. However, this asymmetry tends to disappear gradually by increasing the Finesse. This asymmetry  can be further understood from the fact the $\gamma$ and $\delta$ values may present relevant values, leading to an asymmetry in the generation of the entangled pairs \cite{suppl}.

The observed entanglement is dependent on the temperature, as we observed by exploring this parameter with an analysis frequency of 10 MHz, $\sigma=1.4$ and $\mathcal{F}=15$, as shown in Fig~\ref{fig:Simplectic_Temperaturas}.
Atomic density is a determinant factor for the generation and degradation of entanglement, in a similar behavior to that observed under the Duan criterion analysis.
Individual entanglement against the remaining modes show that internal bands $a_{+\Omega}$ and $b_{-\Omega}$ show a similar behaviour between them, having more entanglement than external bands $a_{-\Omega}$ and $b_{+\Omega}$ at low temperature values.
When temperature increases this behaviour flips and external bands have more entanglement than internal until finally entanglement is lost.

As we have demonstrated, the OPO based on non linear FWM with hot atomic vapor produces entangled twin beams, and an entanglement structure that is consistent with the production of two-mode entangled states involving pairs of sidebands of the signal and idler modes.
This is the first measurement of entanglement, 
in an OPO based on a nonlinear $\chi^{3}$ media, for which the absence of phonon noise comes as an advantage of this gain medium.
The extremely high gain allows for the use of cavities with low finesse, which results in a high escape rate for the produced fields.
On the other hand, the use of a cavity allows the engineering of the generated modes.
With simple improvements, a dedicated setup can be used to produce higher levels of entanglement that may compete with the best values attained by a $\chi^{(2)}$ OPO.
Some limitations, nevertheless, are present.  Atomic density, controlled by the temperature of the $^{85}$Rb vapour cell, is a important parameter that degrades the correlations of the system, until entanglement is lost.
This does not change the fact that the system is a rich and useful tool for the generation of quantum multipartite entangled states that are close to atomic resonances, with applications in many aspects of quantum information.

This work was funded by Grant  No. 2015/18834-0, 2017/27216-4 and 2018/03155-9  S\~ao  Paulo  Research  Foundation (FAPESP), and Grant No. N629091612184 (NICOP/ONR).

%
\end{document}


\preprint{APS/123-QED}

\title{Continuous Variable Entanglement in an Optical Parametric Oscillator Based on a Nondegenerate Four Wave Mixing Process in Hot Alkali Atoms}
\author{A. Monta\~na Guerrero$^1$}
 \email{amontanag@usp.br}
\author{R. Rinc\'on Celis $^1$}
\author{P. Nussenzveig $^1$}
\author{M. Martinelli $^1$}
\author{A. M. Marino $^2$}
\author{H. M. Florez $^1$}
 \email{hans@if.usp.br}
\affiliation{$^1$ Instituto de F\'{\i}sica, Universidade de S\~ao Paulo, 05315-970 S\~ao Paulo, SP-Brazil}
\affiliation{$^2$ Center for Quantum Research and Technology and Homer L. Dodge Department of Physics and Astronomy, The University of Oklahoma, Norman, Oklahoma 73019, USA}

\date{\today}

\pacs{Valid PACS appear here}
\maketitle
\section{Details of the experimental setup}
\begin{figure}[h!]
\begin{overpic}[width=86mm]{Setup1_ver_3.pdf}
\put(0,45){(a)}
\end{overpic}
\begin{overpic}[width=86mm]{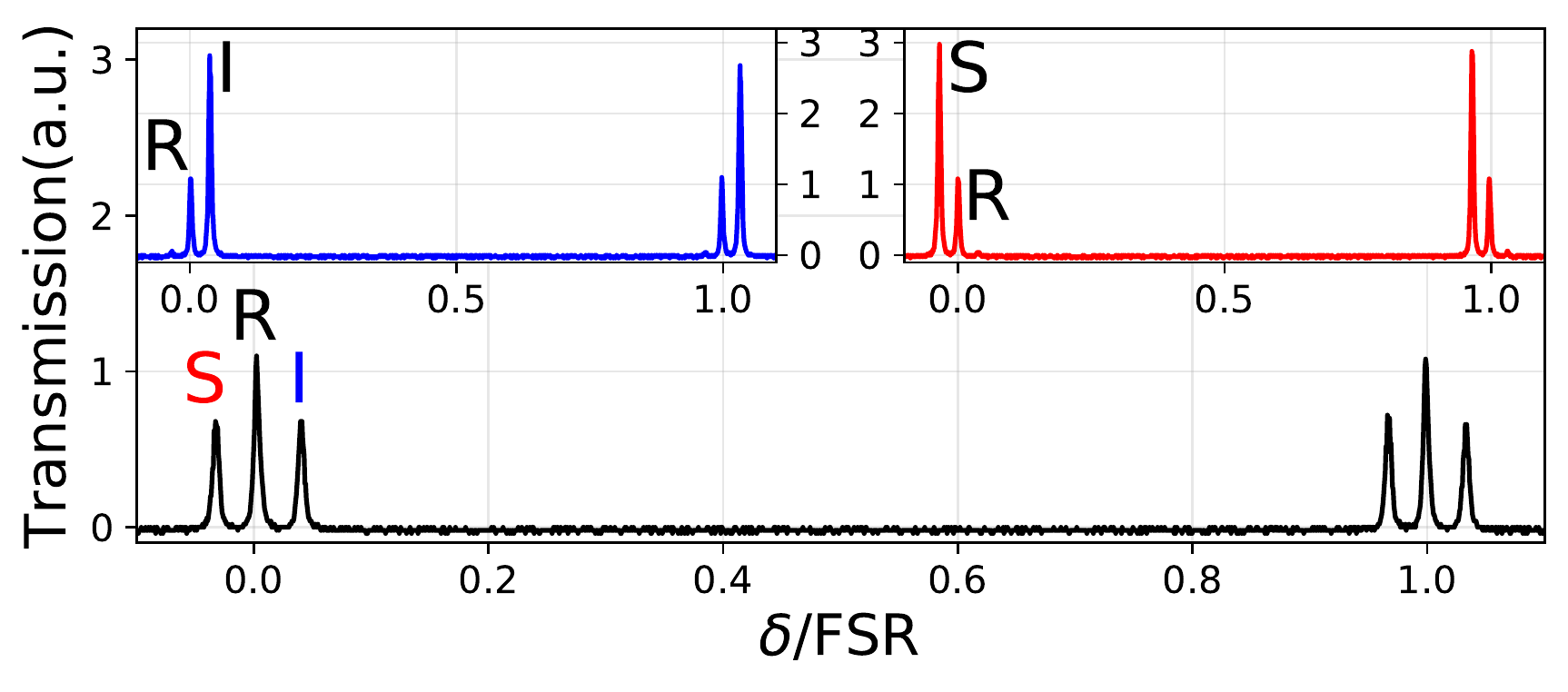}
\put(0,45){(b)}
\end{overpic}
\begin{overpic}[width=86mm]{Carrier.pdf}
\put(0,20){(c)}
\end{overpic}
\caption{(a) Schematic of the experimental setup. PBS: polarising beamsplitter cube, HWP: half-wave plate, UMZ: unbalanced Mach-Zehnder interferometer,M: mirror, FP: confocal Fabry-Perot, FM: flip-mirror, AC: analysis cavity, and PC demodulating chain and data acquisition. (b) Interferometer output analysed with a FP cavity with R, S and I  as reference, signal and idler, respectively. Inset: spectral separation of the signal and idler beams. (c) Representation of the sidebands of mode carriers ``$a$" and ``$b$" and all seven possible bipartitions.}
\label{fig:Setup}
\end{figure}

The experimental setup depicted in Fig.~\ref{fig:Setup} (a), for the generation of entangled beams, is the same as used in \cite{Guerrero20}.
We used a bow-tie cavity  with 4 mirrors with 99.5\% reflectivity, where M$_1$ and M$_2$ have a radius of curvature of 50 cm and M$_3$, M$_4$ are flat mirrors.
Mirror M$_4$ is mounted on a piezoelectric actuator (PZT$_1$) to finely control the cavity length.
The effective cavity size of 74.09 (5) cm corresponds to a free spectral range (FSR) of 404.7 (3) MHz, that is  1/15 of the 6.070 GHz (twice the hyperfine splitting of the ground state of $^{85}$Rb).
The choice of FSR ensures doubly resonant operation and ensures that when the resonant condition is satisfied for both signal and idler modes the pump mode will be out of resonance.
The value of the two waist radius are 316 (3) $\micro $m and 193 (5) $\micro$m for the longer and shorter arms, respectively.
In the waist of the longer arm we place a 3 cm long vapour anti-reflection coated cell with isotopically pure $^{85}$Rb, that is kept at 97$^\circ$C for high optical density.The OPO resonance is controlled by a dither and lock technique.

The pump beam is colinear with the cavity mode and is injected with polarising beam splitter PBS$_1$ and removed with  PBS$_2$.
After PBS$_2$ we use a half wave-plate (HWP) and PBS$_3$ in order to control the output coupling of the cavity, whose finesse can be set from 5 to 30 by changing the orientation of the HWP.
The pump beam is generated with a titanium sapphire laser tuned to the D1(@795nm) line of Rb and is locked to the blue of the $5^2S_{1/2}~F=2\rightarrow 5^2P_{1/2}~F=3$ transition, with a detuning $\Delta_1 =$ 0.82 GHz.

The locking of the laser line is performed by shifting the frequency of a sample of the laser with an acousto-optic modulator (AOM), driven by a frequency of 250 MHz, on a four pass scheme.
The shifted field is sent to a saturated absorption spectroscopy which provides the error signal for the laser frequency stabilization.

We used an unbalanced Mach-Zehnder (UMZ) with a difference of optical path of $\sim$25 mm between both arms, to spectrally discriminated the twin beams generated with the same output  polarization.
The phase shift of the longer arm is adjusted by displacing one of the mirrors with PZT$_{2}$.
The UMZ interferometer gives us ~98\% separation efficiency for the two modes generated by the OPO.
This can be  verified by using a confocal Fabry-Perot cavity (FP) with a 1.5 GHz FSR (see inset Fig. \ref{fig:Setup}(b)).
Taking an injected reference field (R) from the pump beam, we adjust the length of the long path for a constructive interference of the signal (S) or the idler field (I), while verifying that their frequencies correspond to those of the hyperfine splitting.

After the UMZ, we send each beam ($a$ and $b$) to an analysis cavity (AC$_{1,2}$). These bow-tie cavities have a bandwidth of 3.2 MHz, and the fraction of the reflected power at exact resonance is $d=0.38(1)$ for cavity AC$_1$ and $d=0.47(1)$ for cavity AC$_2$.
The bow tie configuration allows a mode-coupling in excess of 99\%. 

Each reflected beam is sent to  a balanced detection, with a pair of associated photodetectors measuring the output of a 50/50 beam splitter.
The advantage of this configuration of detectors is that it enables us to recover, for each measurement, the shot noise level of the incoming beam from the subtraction of the photocurrents, ensuring a precise calibration of the noise level and verification of the quality of the measurement.
Their sum returns the desired quadrature to be measured.

The photocurrent measured on each of the four detectors is splitted in two and demodulated using on-quadrature signals. The demodulation outputs for each detector are named $I_{cos}$ and $I_{sin}$.
We process this signal with on a computer  after acquisition, using and analog-to-digital converter card with an acquisition rate of 600 kHz.
Measurements are performed while scanning the cavity length using the PZT actuators.
During the scan time of 750 ms the cavity resonance crosses over the frequency of the intense fields.

\section{Resonator Detection}
The field transformation given by an optical resonator onto the annihilation operator after reflection from the coupling mirror is $\hat{a} \rightarrow r(\Delta) \hat{a}+t(\Delta)\hat{v}$,
where resonator amplitude reflection  $r(\Delta)$ of  mode $\hat{a}$ and amplitude transmission $t(\Delta)$ of vacuum mode $\hat{v}$ are given by the expressions $r(\Delta)=-(\sqrt{d}+2i\Delta)/(1-2i\Delta)$ and $t(\Delta)=\sqrt{1-r^2(\Delta)}.$
The reflection is calculated under the assumption of high finesse of the analysis cavity  with $\Delta=(\omega-\omega_c)/\delta \nu_{ac}$ representing the detuning between the longitudinal mode frequency $\omega$ and the resonator frequency $\omega_c$ normalized to the resonator bandwidth $\delta \nu_{ac}$. The fraction of reflected light at exact resonance is given by $d=|r(0)|^2$.
The normalized Fourier transform of the photocurrent $I_\Omega=\int I(t) e^{i\Omega t}dt$ can be related to the creation and annihilation operators of the side bands modes as 
$I_\Omega=e^{-i\theta}\hat{a}_\Omega+e^{i\theta}\hat{a}^\dagger_{-\Omega}$ \cite{Barbosa13a}.
The phase relation between the carrier and the sidebands, expressed by $\theta$, can be manipulated by the cavity reflection as $e^{i\theta}=r(\Delta)/|r(\Delta)|$.
Evaluation of the contribution of this phase shift can be performed using the auxiliary functions $g_\pm=X_\pm+iY_\pm$ that in terms of cavity parameters are given by
\begin{eqnarray}
g_+&=&\frac{1}{2} \bigg( \frac{r^*(\Delta)}{|r(\Delta)|}r(\Delta+\Omega)
+ \frac{r(\Delta)}{|r(\Delta)|}r^*(\Delta-\Omega) \bigg),\\
g_-&=&\frac{i}{2} \bigg( \frac{r^*(\Delta)}{|r(\Delta)|}r(\Delta+\Omega)
- \frac{r(\Delta)}{|r(\Delta)|}r^*(\Delta-\Omega) \bigg).
\end{eqnarray}
When considering the additional vacuum fluctuations brought by the cavity losses, we get the expression
$I_\Omega=(I_{cos}+i I_{sin})/\sqrt{2}+I_{vacuum}$ with $[I_{cos},I_{sin}]=0$.
Notice that the cavity losses are not a problem in this case, but rather a necessary condition to recover the energy unbalance between the sidebands of a single beam \cite{Villar20}.
\begin{figure}[h]
    \centering
    \includegraphics[width=86mm]{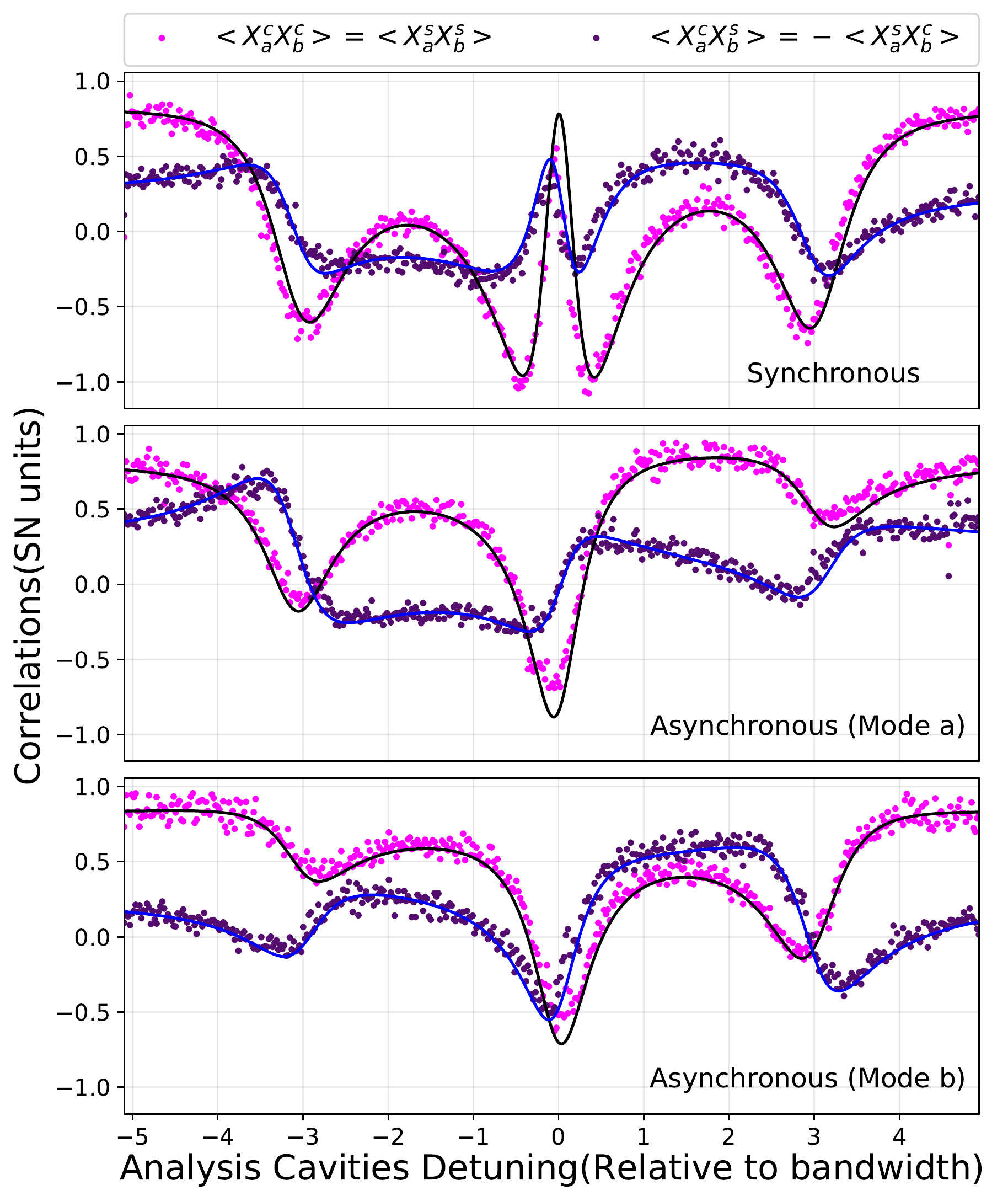}
    \caption{Correlations between the generalized quadratures $X_a$ and $X_b$.
    Superscripts c and s represent cosine and sine demodulation of photocurrents.}
    \label{fig:Corr_XiXj_cs}
\end{figure}
For a full description of the photocurrent $I_\Omega$ at a given analysis frequency $\Omega$ we need to measure two photocurrents, $I_{cos}$ and $I_{sin}$, that are in quadrature.
This is the reason why we split each photocurrent and demodulated with on-quadrature signals.
These operators depend on linear combinations of symmetric and antisymmetric side band modes,
\begin{eqnarray}
I_{cos}=X_+\hat{p}_\textup{s}+X_-\hat{q}_\textup{s}+Y_-\hat{p}_\textup{a}-Y_+\hat{q}_\textup{a}\\
I_{sin}=Y_+\hat{p}_\textup{s}+Y_-\hat{q}_\textup{s}-X_-\hat{p}_\textup{a}+X_+\hat{q}_\textup{a}
\end{eqnarray}
with $\hat{p}_{(\textup{s},\textup{a})}=(\hat{p}_{\Omega} \pm \hat{p}_{-\Omega})/\sqrt{2}$ and  $\hat{q}_{(\textup{s},\textup{a})}=(\hat{q}_{\Omega} \pm \hat{q}_{-\Omega})/\sqrt{2}$, where $\hat{p}_{\Omega}=(\hat{a}_{\Omega}+\hat{a}^{\dagger}_{\Omega})$ and $\hat{q}_{\Omega}=-i(\hat{a}_{\Omega}-\hat{a}^{\dagger}_{\Omega})$ are linear combinations of sideband modes $\hat{a}_{\pm\Omega}$ ($\hat{b}_{\pm\Omega}$) as depicted in Fig.~\ref{fig:Setup}(c).
The photocurrent spectral noise power of Resonator Detection (RD) for each beam is  
\begin{eqnarray}
    S_{RD}=\langle I_\Omega I_{-\Omega} \rangle=\frac{1}{2}\Delta ^2 I_{cos}+\frac{1}{2}\Delta ^2 I_{sin}+\Delta ^2 I_{v},
\end{eqnarray}
with stationary conditions $\Delta ^2 I_{cos}=\Delta ^2 I_{sin}$, that can be reorganized in terms of the coefficients of the covariance matrix $V_{(\textup{s}/\textup{a})}^{(12)}$ (Eq. 3 in the main text)
\begin{eqnarray}\label{eq:Variance}
    S_{RD}(\Delta)=c_\alpha \alpha+c_\beta \beta+c_\gamma \gamma +c_ \delta \delta +c_v,
\end{eqnarray}
where $c_\alpha=\big| g_+ \big|^2$, $c_\beta=\big| g_- \big|^2$, $c_\gamma+i c_\delta = 2g^\ast_+ g_-$ and $c_v=1-c_\alpha-c_\beta$. The correlations between two modes are given by $S_{RD}=\big\langle I^{(1)}_\Omega I^{(2)}_{-\Omega}\big\rangle$
\begin{eqnarray}
\Re{\big\langle I^{(1)}_\Omega I^{(2)}_{-\Omega}\big\rangle}=\frac{1}{2}\big\langle I^{(1)}_{cos} I^{(2)}_{cos}\big\rangle+\frac{1}{2}\big\langle I^{(1)}_{sin} I^{(2)}_{sin}\big\rangle
\\
\Im{\big\langle I^{(1)}_\Omega I^{(2)}_{-\Omega}\big\rangle}=\frac{1}{2}\big\langle I^{(1)}_{sin} I^{(2)}_{cos}\big\rangle-\frac{1}{2}\big\langle I^{(1)}_{cos} I^{(2)}_{sin}\big\rangle
\end{eqnarray}
\begin{figure}
    \centering
    \includegraphics[width=86mm]{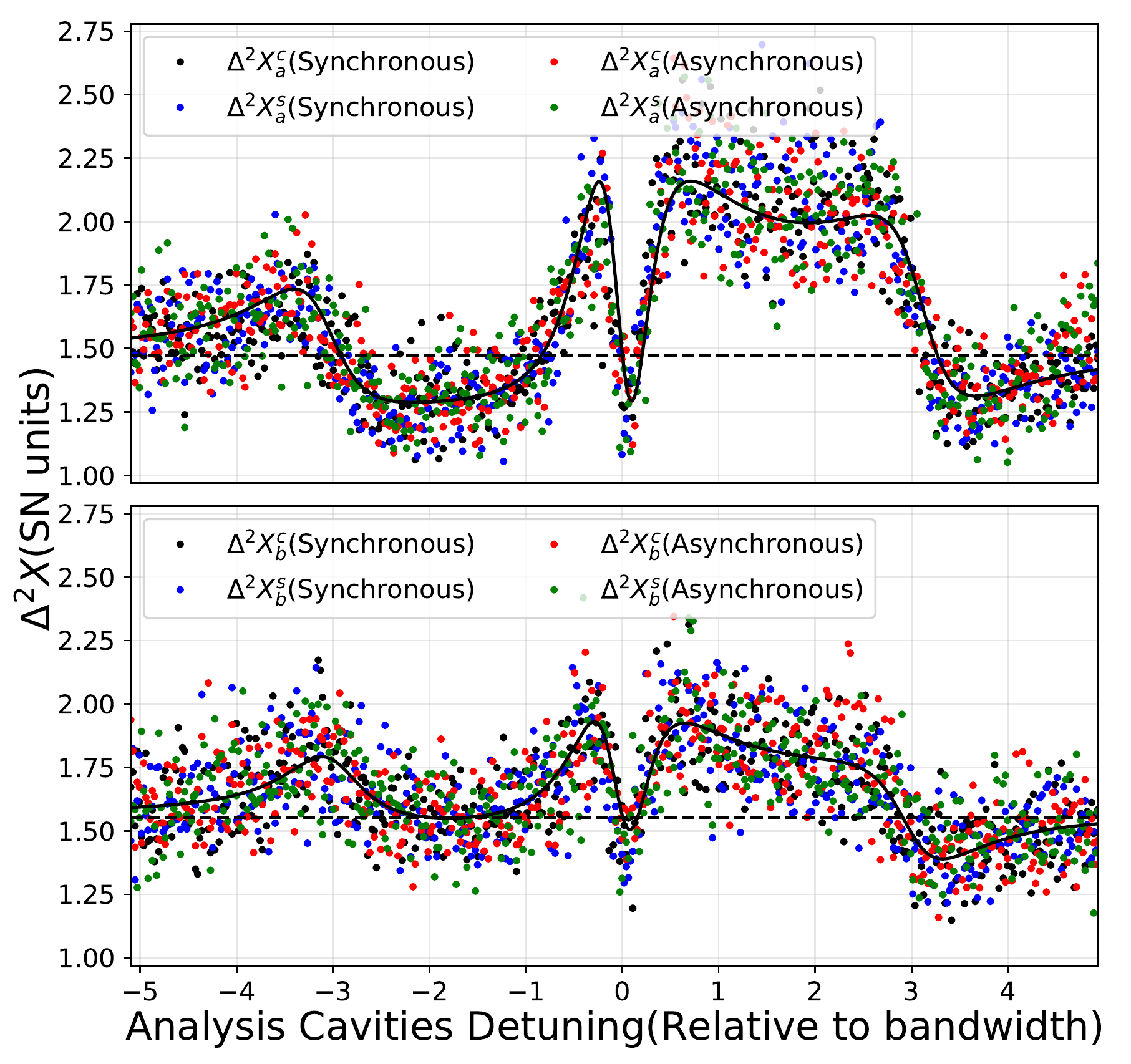}
    \caption{Variance of the generalized quadrature $X_a$ ($X_b$) for mode $a$($b$). We have four redundant data set for each mode. Dotted line represent a cavity out of resonance. Superscripts c and s represent cosine and sine demodulation of photocurrents. }
    \label{fig:Var_Xi_cs}
\end{figure}
with stationarity conditions $\langle I^{(1)}_{cos} I^{(2)}_{cos}\rangle=\langle I^{(1)}_{sin} I^{(2)}_{sin}\rangle$ and $\langle I^{(1)}_{sin}I^{(2)}_{cos}\rangle=-\langle I^{(1)}_{cos} I^{(2)}_{sin}\rangle$. These terms can be reorganized as well, resulting in
\begin{eqnarray}\label{eq:Correlations}
\Re{\big\langle I^{(1)}_\Omega I^{(2)}_{-\Omega}\big\rangle}=c_\mu \mu +c_\eta \eta +c_\epsilon \epsilon+c_\kappa \kappa \nonumber \\
+c_\xi \xi +c_\lambda \lambda+c_\nu \nu+c_\tau \tau,\\
\Im{\big\langle I^{(1)}_\Omega I^{(2)}_{-\Omega}\big\rangle}=c_\mu \eta-c_\eta \mu +c_\epsilon \kappa-c_\kappa \epsilon \nonumber \\\label{eq:correlationsIm}
+c_\xi \lambda -c_\lambda \xi+c_\nu \tau-c_\tau \nu,
\end{eqnarray}
where $c_\mu+i c_\eta = g^\ast_{+_1} g_{+_2}$, $c_\epsilon+i c_\kappa = g^\ast_{+_1} g_{-_2}$, $c_\xi+ c_\lambda = g^\ast_{-_1} g_{+_2}$ and $c_\nu+i c_\tau = g^\ast_{-_1} g_{-_2}$.
Notice that all $c_i$ functions depend only on known cavity parameters, and the DC part of the photocurrent gives a precise evaluation of the detuning scale, resonance position and intensity attenuation $d$ that is fed as fixed parameters in the analysis of the fluctuating (AC) part.
We have three stages of measurement as we can see in Fig~\ref{fig:Corr_XiXj_cs}: First, when we have synchronous scanning of both analysis cavities AC$_1$ and AC$_2$ all eight functions $c_i=c_\mu,c_\eta,..$ are different form zero.
Secondly, in the case when we only scan AC$_1$ and put AC$_2$ out off resonance, functions $c_\epsilon,c_\kappa,c_\nu,c_\tau$ are null.
Finally, when we only scan AC$_2$ and put AC$_1$ put out off resonance, functions $c_\xi,c_\lambda,c_\nu,c_\tau$ are null.
This provides us redundancy and precision at the moment of fitting the parameters $i=\mu,\eta,\cdots$.
The variance of a single beam returns the tomography of that single pair of modes, associated to $\alpha$, $\beta$, $\delta$, and $\gamma$ . From Fig. \ref{fig:Var_Xi_cs} we can see that these curves remain compatible during the three acquisitions shown in Fig. \ref{fig:Corr_XiXj_cs}.
We use the package LMFIT (Non-Linear Least-Squares Minimization and Curve-Fitting for Python) to fit our data to Eqs. \ref{eq:Variance},  \ref{eq:Correlations} and \ref{eq:correlationsIm} as shown in Figs.~\ref{fig:Corr_XiXj_cs} and \ref{fig:Var_Xi_cs}, respectively.
This give us the full 16 parameters of the covariance matrix in the symmetric/antisymmetric basis.